\documentclass[aps,prl,reprint,twocolumns,notitlepage,superscriptaddress]{revtex4-1}
\usepackage{amssymb,bm}
\usepackage{amsmath,eucal,ulem}%,cancel}
\usepackage[usenames,dvipsnames]{color}
\usepackage{graphicx}
\usepackage{units}
\usepackage{epstopdf}

\begin{document}

\title{Measurement of Anomalous Diffusion Using Recurrent Neural Networks}
\author{Stefano Bo}
\altaffiliation[Present address: ]{Max Planck Institute for the Physics of Complex Systems, N{\"o}thnitzer Str. 38, DE-01187 Dresden, Germany.}
\affiliation{Nordita, Royal Institute of Technology and Stockholm University,
Roslagstullsbacken 23, SE-106 91 Stockholm, Sweden}
\author{Falko Schmidt}
\affiliation{Department of Physics, University of Gothenburg, SE-412 96 Gothenburg, Sweden}

\author{Ralf Eichhorn}
\affiliation{Nordita, Royal Institute of Technology and Stockholm University,
Roslagstullsbacken 23, SE-106 91 Stockholm, Sweden}
\author{Giovanni Volpe}
\affiliation{Department of Physics, University of Gothenburg, SE-412 96 Gothenburg, Sweden}

\begin{abstract}
Anomalous diffusion occurs in many physical and biological phenomena, when the growth of the mean squared displacement (MSD) with time has an exponent different from one.
We show that recurrent neural networks (RNN) can efficiently characterize  anomalous diffusion by determining the exponent from a single short trajectory, outperforming the standard estimation based on the MSD when the available data points are limited, as is often the case in experiments. 
Furthermore, the RNN can handle more complex tasks where there are no standard approaches, such as determining the anomalous diffusion exponent from a trajectory sampled at irregular times, and estimating the switching time and anomalous diffusion exponents of an intermittent system that switches between different kinds of anomalous diffusion.
We validate our method on experimental data obtained from sub-diffusive colloids trapped in speckle light fields and super-diffusive microswimmers.
\end{abstract}

\date{\today}

\maketitle

Anomalous diffusion underlies various physical and biological systems, such as the motion of microscopic particles in a crowded subcellular environment and the active  dynamics of biomolecules in the cytoplasm \cite{Barkai2012, Hofling2013, Metzler2014}.
While normal diffusion is characterized by a linear growth of the mean squared displacement (MSD) with time, anomalous diffusion features a non-linear, power-law growth.
If we consider a microscopic particle  whose position is $X(t)$, its MSD is, in the stationary case, 
\begin{equation}\label{eq:def}
{\rm E} \left[
\left(
X(t+\tau) - X(t)
\right)^2
\right]
=
K_\alpha \tau^\alpha,
\end{equation}
where $\alpha$ is the exponent characterizing the anomalous diffusion and $K_\alpha$ is a generalized diffusion coefficient with dimension $[\mbox{length}^2\,\mbox{time}^{-\alpha}]$.
The exponent $\alpha$ contains crucial information regarding the nature of these systems distinguishing standard diffusion ($\alpha = 1$) from anomalous diffusion ($\alpha < 1$ for sub-diffusion and $\alpha > 1$ for super-diffusion).
Therefore, it is crucial to be able to determine its value from experimental data.
When large datasets are available, the exponent can be straightforwardly fitted from the empirical MSD \cite{Golding2006,Bronstein2009,Weber2010,Jeon2013,Caspi2000}, or using alternative techniques  \cite{Tejedor2010,Burnecki2015,Meroz2015,Makarava2011, Hinsen2016,Krapf2019,Weron2019}.
Most of these methods work under the assumption that the exponent does not change abruptly over the duration of the measurement, and require the particles' trajectory to be sufficiently long and to be sampled at regular time intervals (unless  several trajectories are available for each case).

However, especially in single-molecule studies and in non-equilibrium experiments, the dynamic and unsteady character of the process under study and the variability of the environment restrict the possibility to collect large amounts of data under the exact same conditions \cite{Jeon2010a}. 
Therefore, often one has only access to trajectories that are short (e.g., limited measurement time \cite{Weron2019,Meroz2015,Elf2019}), that are sampled at irregular times (e.g., due to fluorophore blinking \cite{Elf2019}), or whose diffusion properties change over time (e.g., intermittent anomalous diffusion \cite{Akin2016,Sikora2017c,Weron2017}).
In these cases, the standard approaches based on the MSD cannot be straightforwardly employed. 
Instead, suitable approaches need to be developed on a case-by-case basis --- a process that is often time-consuming and subject to user bias.

Recently, data-driven approaches have emerged as an alternative paradigm to analyze experimental data in several branches of physics and biology \cite{Zdeborova2017, Helgadottir2019}.
While standard algorithms require the user to explicitly give rules to process the input data in order to obtain the sought-after result, data-driven algorithms are trained through a large series of input data and the corresponding desired outputs from which they autonomously determine the rules for recognizing patterns.
In this way, data-driven approaches can make  very efficient use of all the information contained in the available data. 
Neural networks are one of the most successful data-driven approaches in estimation and regression tasks due to their great ability   
to automatically learn from data~\cite{Nielsen2015,Lipton2015a}. This feature has been successfully employed in a number of tasks ranging from hand-written digits and image recognition to natural language translation \cite{Wu2016}.
Therefore, neural networks   ideally complement standard techniques to perform inference in cases for which no standard algorithmic procedures are available.
In fact, some seminal works have already applied machine-learning techniques to determine the properties of anomalous diffusion with a focus on identifying its underlying mechanisms  \cite{Wagner2017, Munoz-Gil2019, Granik}. 
We remark that, similarly to other advanced machine learning techniques, neural networks often operate as black boxes and therefore should be applied carefully to new experimental data and situations, always testing and benchmarking their performance against established techniques.

In this contribution, we show that recurrent neural networks (RNN) can successfully be employed to characterize anomalous diffusion.
While RNN perform equally well as MSD approaches when characterizing the anomalous diffusion from sufficiently long, regularly sampled and stationary time series, RNN marginally outperform MSD approaches when the available data points are limited, as is often the case in experiments.
More importantly, RNN can also straightforwardly deal with more complex cases for which there are no standard approaches: when trajectories are sampled at irregular sampling times, and when the system features an intermittent behavior.
We validate the use of RNN on experimental data obtained from colloids sub-diffusing in a speckle light field \cite{Volpe2014, volpe2014speckle} and microswimmers super-diffusing when illuminated \cite{Schmidt2019}.

\begin{figure}[t!]
\includegraphics{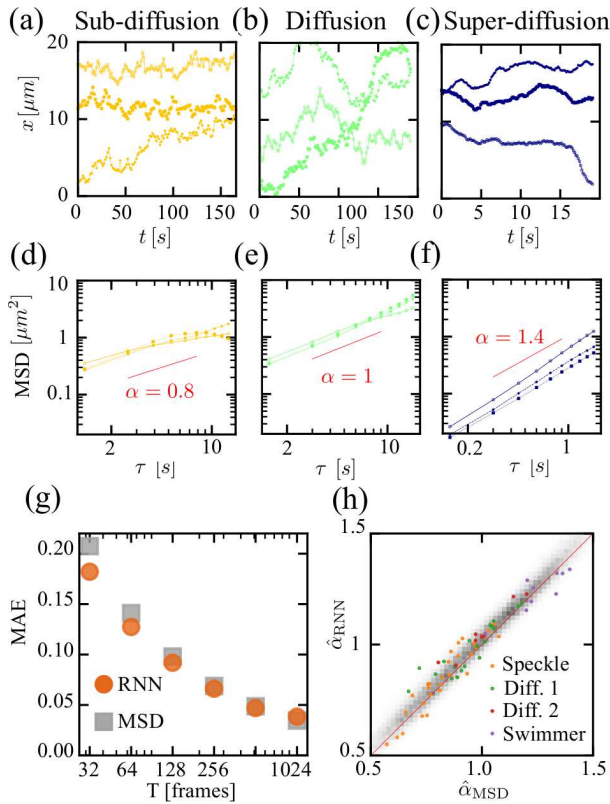}
\caption{Measurement of anomalous diffusion with recurrent neural networks (RNN).
(a-c) Experimental trajectories of a particle undergoing (a) sub-diffusion (motion in a speckle light field), (b) normal diffusion, and (c) super-diffusion (light-activated microswimmer).
(d-f) Corresponding MSDs.
(g) Mean absolute error (MAE) of the exponent inferred using the standard time-averaged MSD (gray squares) and the RNN (orange circles) as a function of the trajectory length. The performances are tested on 185000 simulated trajectories undergoing fractional Brownian motion with $\alpha$ uniformly sampled between $[0.5, 1.5]$. 
(h) Exponents estimated by the RNN vs. those estimated by the MSD.
The gray background represent a density plot of exponents obtained from simulated trajectories  and the colored points represent exponents obtained from experimental data: orange for the sub-diffusive colloids in a speckle light field, green for the same colloids freely diffusing,  purple for the super-diffusive microswimmers, and red for the inactive microswimmers that  diffuse normally.
}\label{fig:1} 
\end{figure}

Figs.~\ref{fig:1}a-c show some examples of experimental trajectories ($128$ measurement points each, see Supplemental Material \cite{SM}) corresponding to a colloidal microsphere (${\rm SiO}_2$, radius $R=2.5\,{\rm \mu m}$) undergoing sub-diffusion in a speckle light field \cite{Volpe2014} (Fig.~\ref{fig:1}a), the same colloid without the speckle light field  normally diffusing (Fig.~\ref{fig:1}b), and  a microswimmer (${\rm SiO}_2$ microsphere with iron-oxide inclusions, $R=0.49\,{\rm \mu m}$) in a critical mixture that super-diffuses when illuminated by light \cite{Schmidt2019} (Fig.~\ref{fig:1}c).
The time-averaged-MSD is calculated from each these trajectories as 
\begin{equation}
\mbox{MSD}(\tau)=\frac{1}{T/\delta t-j+1}
\sum_{i=0}^{T/\delta t-j} 
\left(
X_{i+j} - X_i
\right)^2,
\end{equation}
where the discrete measurements $X_i=X(i\delta t)$ are taken at intervals $\delta t$, and the time lag is given by $\tau=j\delta t$.
The corresponding MSDs are plotted by colored lines in log-log scale in Figs.~\ref{fig:1}d-f and the value of the exponent is obtained from linear interpolation. 
Clearly, for these short trajectories a precise estimation of the exponent is challenging \cite{Jeon2010a}:
there is some arbitrariness in what segments of the trajectories or of the MSD plots to use for the fitting;
and the choice depends on the specific $\alpha$, on the measurement noise, and on the length of the trajectory, so that additional {\it a priori} knowledge about the system is required \cite{Kepten2013, Kepten2015, Lanoiselee2018}.

We propose a new method based on RNN to determine $\alpha$ directly from the single trajectories. 
RNN are ideal to deal with time sequences because they process the input data sequence iteratively and, therefore, explicitly model the sequentiality of the input data \cite{Hochreiter1997,Lipton2015a}.
In fact, differently from other neural network architectures that process the input data at once (e.g., dense and convolutional neural networks), RNN loop over the input data sequence, keeping an internal model of the information they are processing, built from past information and constantly updated as new information arrives \cite{Hochreiter1997,Lipton2015a}.
Thanks to their recurrent nature, RNN typically require fewer layers to perform a given task than alternative neural network architectures; for example, the neural network that currently powers the Google Translate algorithm is a stack of just seven large ``long short term memory'' (LSTM) layers \cite{Wu2016}.
We employ a RNN constituted of two LSTM layers with states of dimension $64$ and $16$, respectively, and a densely connected output layer, which provides the estimate of the exponent $\hat{\alpha}$ \cite{SM}.
We have implemented this neural network using the Python-based Keras library \cite{chollet2015keras} with a TensorFlow backend \cite{Abadi2016} because of their broad adoption in research and industry; nevertheless, we remark that the approach we propose is independent of the  framework used for its implementation.

Once the network architecture is defined, we need to train it on a set of single trajectories for which we know the ground-truth values of $\alpha$. For each trajectory containing $T$ measurement points, the input data to the network is  a $2\times T$-dimensional array containing  position and  time  for each measurement point
$\left[ \left( x_1,t_1 \right), \, \left( x_2,t_2 \right), \ldots, \left( x_T,t_T \right) \right]$
 (suitably normalized so that the position's average and standard deviation of a trajectory are, respectively, $0$ and $1$ and the rescaled measurement times are  between $0$ and $1$, as discussed in the Supplemental Material~\cite{SM}).
In each training step, the neural network is tasked with predicting the exponents corresponding to each trajectory from a batch of the training set; its predictions are then compared to the ground-truth values of the exponents; and the prediction errors are finally used to adjust the trainable parameters of the neural network using a back-propagation algorithm \cite{Hochreiter1997,Lipton2015a}. 
  The training of a neural network is notoriously data intensive, requiring in our case several hundreds of thousand to millions trajectories. In order to have enough trajectories and to accurately know the ground-truth values of the corresponding exponents, we simulate the trajectories.
There are several models and mechanisms that can give rise to anomalous diffusion dynamics \cite{Metzler2014} and several methods to identify such models from data (e.g.,~\cite{Meroz2015,Granik,Magdziarz2009,Burnecki2012,Thapa2018}).
We choose to  train using fractional Brownian motion (fBm)%, which generates anomalous diffusion dynamics 
~\cite{Mandelbrot1968}, which
% has been shown to describe several processes in biophysics \cite{Burnecki2012};
%fBm
 is defined as a continuous-time Gaussian process ($B_\alpha(t)$) with zero mean and correlated increments that give rise to the covariance function
\begin{equation}
E\left[B_\alpha(t)B_\alpha(s)\right]=\frac{1}{2}\left(|t|^{\alpha}+|s|^{\alpha}-|t-s|^{\alpha}\right)\label{eq:def},
\end{equation}
where $\alpha$ is the exponent with which the mean squared displacement grows (Eq.~\ref{eq:def}). % and is equal to twice the Hurst parameter  \cite{Mandelbrot1968}.
We simulate the trajectories using the Davies-Harte and the Hosking algorithm \cite{Dieker2004} implemented in a Python library \cite{Flynn}.

To assess the performance of the RNN, we test it on independently simulated trajectories with $\alpha$ uniformly sampled in $[0.5, 1.5]$ against %the performance of 
the MSD, 
because this is the most widespread and easy-to-use method in soft-matter and biophysics experiments.
We linearly fit the time-averaged MSD for $\tau=1,...,5$, which delivers a
good performance for fractional Brownian motion in cases without measurement noise.
In case of the RNN, we train a different network for each of the different trajectories' lengths we consider.
Fig.~\ref{fig:1}g shows the mean absolute error (MAE) for the two  methods as a function of the trajectory's length.
For long trajectories, both methods perform similarly well with MAE for RNN (MSD) 
$0.038$ ($0.035$), $0.047$ ($0.049$) and $0.066$ ($0.069$) for trajectories with $1024$, $512$ and $256$ samples, respectively. For shorter trajectories, which are known to be problematic for MSD-based methods \cite{Jeon2013, Kepten2015}, the RNN performs slightly better, achieving MAE $0.092$ (vs. $0.098$ for the MSD), $0.127$ (vs. $0.141$) and $0.182$ (vs. $0.207$) for $128$, $64$ and $32$ samples, respectively.
This demonstrates that the RNN is able to extract  information from the trajectories that is not used by the MSD. We remark that there is some variability in the performance of the networks across different trainings and that, focusing on a specific trajectory length, it is possible to further improve the predictions by fine-tuning the training and by pooling the predictions of different networks.
In any case, the predictions made with the RNN and the MSD are strongly correlated, as can be seen in Fig.~\ref{fig:1}h where the estimations made using the RNN are plotted against the ones for the MSD for simulated trajectories of length $128$. 
Importantly, even though the RNN is trained on a specific model for anomalous diffusion (fBM), it is able to generalize and to correctly analyze also experimental data for which we do not know the precise mechanism underlying the anomalous diffusion behavior.
The colored points in Fig.~\ref{fig:1}g represent the estimations made using the RNN plotted against the ones for the MSD for the experimental data corresponding to sub-diffusive particles moving in a speckle light field (orange points), to diffusive Brownian particles (green and red points), and to super-diffusive microswimmers (purple points).
The RNN and MSD estimations are correlated in a similar way as for the data generated from simulations using a fBM model, providing strong evidence for the experimental reliability of the RNN method even when the underlying microscopic dynamics for the anomalous diffusion are other than fBM.

In the next step, we show that RNN can be used to determine $\alpha$ in two cases where a straightforward computation of the MSD becomes challenging:
trajectories are sampled at irregular  times, and a system featuring an intermittent behavior.

\begin{figure}[t!]
\includegraphics[width=\columnwidth]{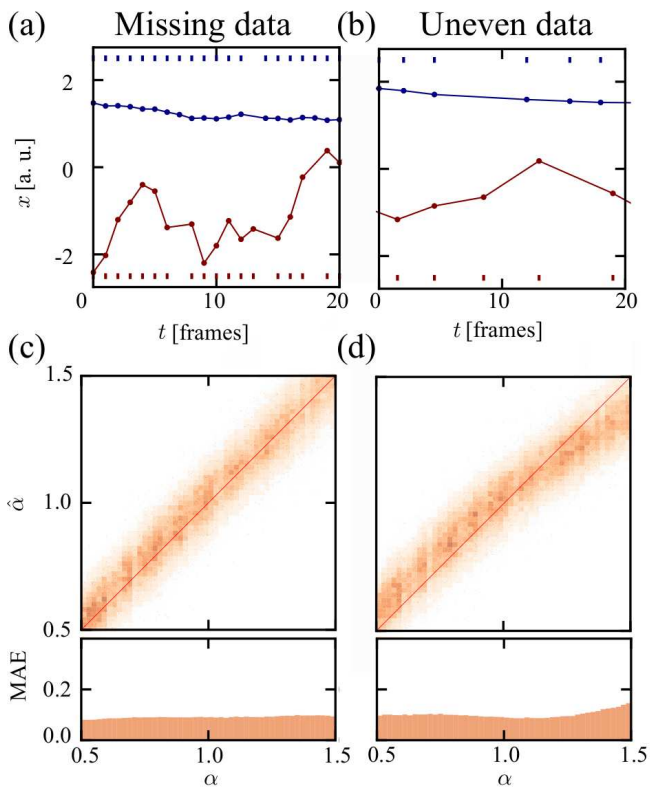}
\caption{Measurement of anomalous diffusion in irregularly sampled trajectories.
Often a trajectory is sampled irregularly, either (a) because some data points are missing (``missing data'', here $12.5\%$ data points are missing), or (b) because data points are sampled at random times (``uneven data'', here according to a geometric distribution). For each case, two trajectories with different $\alpha$ are shown.
(c-d) Estimated exponent $\hat{\alpha}$ as a function of the actual exponent $\alpha$ for the two cases using simulated trajectories with $128$ frames. 
The lower panels show the MAE as a function of $\alpha$.
The MAE averaged over all $\alpha$ is $0.091$ in (c) and  $0.101$ in (d). There exists a systematic bias in the more challenging ``uneven data'' case, visible in (d) for large $\alpha$.}
\label{fig:2}
\end{figure}

The first situation is motivated by the fact that,
in several experimental settings, it is not possible to record the trajectories at equally spaced time intervals. 
For example, the fluorescent biomarkers commonly employed for tracking biomolecules are subject to blinking so that some portion of a trajectory might be missing \cite{Elf2019}.
In general, tracking algorithms might miss some frames, especially in noisy and challenging experimental conditions, leading to trajectories with missing data points.
No standard technique exists to deal with these cases for single trajectories.
Here, we test the RNN, trained in Fig.~\ref{fig:1} on trajectories sampled at regular times, on trajectories sampled at irregular times.
We consider two scenarios: 
(a) a fraction of the regularly recorded data is missing (``missing data'' scenario, Fig.~\ref{fig:2}a); 
(b) the data points are sampled at random times (``uneven data'' scenario, Fig.~\ref{fig:2}b).
As shown in Figs.~\ref{fig:2}c-d, the RNN is in fact able to generalize to these cases.
In particular, for the ``missing data'' scenario with $12.5\%$ data points randomly missing, the performance of the network is unaffected as long as the same number of data points as in the training set (in this case $128$) is fed into the RNN (Fig.~\ref{fig:2}c).
For case (b), with measurement times geometrically distributed so that on average $1$ frame every $8$ contains a signal, the RNN  provides reasonable predictions, which however are slightly biased and  tend to underestimate large exponents (Fig.~\ref{fig:2}d).
For more accurate predictions, one can retrain the RNN on irregularly sampled simulated data and significantly improve its performance \cite{SM}.

 \begin{figure}[t!]
 \includegraphics[trim=0cm 0.cm 0.cm 0.cm, clip=true, width=\columnwidth]{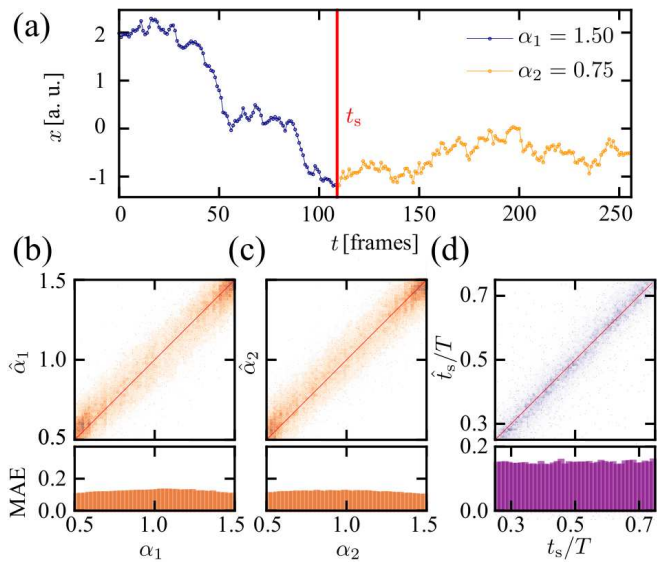}
\caption{Measurement of the switch between two anomalous diffusion behaviors.
(a) Simulated trajectory of a particle whose exponent switches from $\alpha_1 = 1.50$ to $\alpha_2 = 0.75$ at time $t_{\rm s} = 108$.
Estimation by a RNN of (b) 
$\hat{\alpha}_1$, (c) $\hat{\alpha}_2$, and (d) $\hat{t}_{\rm s}/T$ as a function of the respective ground-truth values, for a test data set where $|\Delta \alpha |>0.25$, $t_{\rm s} \in [0.25 T, 0.75 T]$, and $T=256$.
}
\label{fig:3}
\end{figure}

\begin{figure}[t!]
\includegraphics[width=\columnwidth]{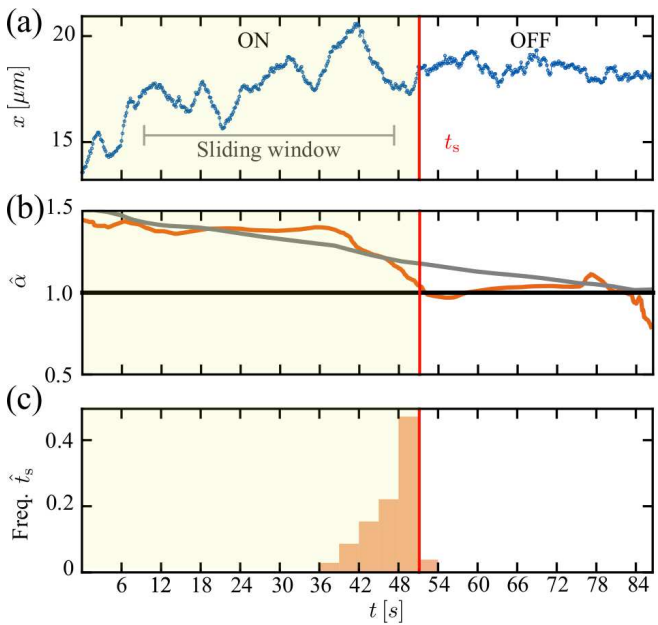}
\caption{Determination of anomalous diffusion exponents and switching time in an experimental time series by sliding a window containing $256$ measurement points ($T= 38.3$\,{\rm s}).
(a) Trajectory of a microswimmers activated by light ($\alpha \simeq1.4$); at $t=51.3\,{\rm s}$ the light is switched off and the microswimmers becomes a passive Brownian particle ($\alpha = 1.0)$. 
(b) Orange curve: exponent $\hat{\alpha}$ estimated by averaging the predictions of the RNN for each sliding window.
For reference, the gray curve reports the exponent $\hat{\alpha}$ estimated by averaging the predictions of the MSD  for the same sliding windows. % sliding a window of containing $256$ measurement points.}
(c) Histogram of the switching times $\hat{t}_{\rm s}$  estimated by the RNN. Each prediction is obtained from   a different starting point of the  sliding window %with a different starting point 
and the histogram is built  from  reliable windows where $|\Delta\hat{\alpha}|>0.25$ and the estimated change point is far from the boundaries of the window $\hat{t}_{\rm s} \in [0.25 T, 0.75 T]$.
}
\label{fig:4}
\end{figure}

As the second situation, we consider systems featuring intermittent behavior, where the particle diffusion switches between different behaviors characterized by different $\alpha$.
Such behavior occurs, for instance, when particles are transiently trapped such as in sodium channels \cite{Akin2016} or when self-propulsion is switched on and off \cite{Schmidt2019}.
Relying on traditional MSD measurements, one would first need to detect the change in behavior (e.g., using change-point analysis techniques \cite{D.Montiel2006}) and successively to estimate the exponents of the two sub-trajectories. 
This is a challenging procedure, which has been attempted only recently for trajectories switching between sub-diffusive and super-diffusive dynamics \cite{Sikora2017c, Sikora2018}.
We employ a modified version of the RNN discussed above to determine simultaneously the exponent before switching $\hat{\alpha}_1$, the exponent after switching $\hat{\alpha}_2$, and the switching time $\hat{t}_{\rm s}$ from the acquired trajectory.
Specifically, we use a network with the same architecture as before but with $5$ output neurons that estimate $\hat{\alpha}_1$, $\hat{\alpha}_2$, $\Delta \hat{\alpha} = \hat{\alpha}_2-\hat{\alpha}_1$, $\sin\left(2\pi\hat{t}_{\rm s}/T\right)$, and $\cos\left(2\pi\hat{t}_{\rm s}/T\right)$ (see \cite{SM}). 
We train this RNN on a set of $1.6$-million simulated trajectories where a change in exponent occurs randomly with a uniform distribution at time $t_{\rm s}$ (see \cite{SM}).
Figs.~\ref{fig:3}b-d shows the performance of the estimations of $\hat{\alpha}_1$ (Fig.~\ref{fig:3}b), $\hat{\alpha}_2$ (Fig.~\ref{fig:3}c), and $\hat{t}_{\rm s}$ (Fig.~\ref{fig:3}d), when the change in $\alpha$ is not too small ($|\Delta \alpha|>0.25$) and the switch occurs around the middle of the trajectory ($t_{\rm s} \in [0.25 T, 0.75 T]=[64, 192]$). Under these conditions,
the performance in estimating $\hat{\alpha}_1$ (Fig.~\ref{fig:3}b, MAE $0.116$) and $\hat{\alpha}_2$ (Fig.~\ref{fig:3}c, MAE $0.112$) is comparable to the case of constant $\alpha$ reported in Fig.~\ref{fig:1}.
The switching-time estimation can be challenging when the change of the exponent is small or occurs very early/late. For $\hat{t}_{\rm s}/T$ we have a MAE $0.148$ as illustrated in  Fig.~\ref{fig:3}d.

In Fig.~\ref{fig:4}, we illustrate the power of the neural-network approach using an experimental trajectory. 
We consider a microswimmer, which undergoes super-diffusion ($\alpha \simeq 1.4$) when illuminated by light \cite{Schmidt2019}, and becomes diffusive when the light is turned off ($\alpha = 1.0$).
Fig.~\ref{fig:4}a shows the corresponding trajectory with a switching time at $t=51.3\,{\rm s}$.
We measure this switch using the RNN with a sliding window containing $256$ measurement points ($38.3 \,{\rm s}$). 
The estimated  exponent (averaged over the various sliding windows) is shown in Fig.~\ref{fig:4}b, where one can see that there is a clear shift from $\hat{\alpha}\simeq1.4$ to $\hat{\alpha}=1.0$ around $t=50\,{\rm s}$.
As a reference, we show the prediction from the MSD (gray curve in Fig.~\ref{fig:4}b) obtained by averaging the exponents inferred by a sliding window. One can see that, in this case, the transition between the high and low exponent is smoothed out and takes place in a longer time interval. One could try to alleviate the issue by choosing shorter window sizes but this would come at the cost of a noisier estimation.
The histogram of the  switching times predicted by the RNN in the different windows is shown in Fig.~\ref{fig:4}c, where it can be seen that the network correctly determines the switching time.

In conclusion, we introduced a new method for the estimation of the exponent from single trajectories in anomalous diffusion systems based on RNN.
We have shown that it can be straightforwardly applied to more complex situations, where standard approaches are lacking.  Our method then emerges as a promising tool for the analysis of single trajectories with irregular measurements and intermittent behaviors.
We remark that our analysis has been limited to the case in which the observed time series can be described by a single exponent on the observation time scales (or a distinct switch between two exponents). 
In several systems, the MSD smoothly interpolates between different linear (in the log-log plot) regimes on different time scales (see, e.g., \cite{Bronstein2009,Jeon2013}).
The approach we propose here should not be directly applied to analyze such time series on time scales where the transition between different diffusive regimes occurs, but has to be separately applied to the different linear regimes.
In general, when dealing with completely unseen data, before proceeding to a deeper analysis, it is advisable to  benchmark the preliminary predictions of the RNN against the ones of the MSD. 
As future work, it will be interesting to test the inference of the RNN method trained on fBM simulated data on data obtained from different anomalous diffusion models, such as, for instance, continuous time random walks. It would also be possible to train the RNN using
simulated data not generated from fBM. Along these lines, it would be interesting to consider higher-order moments, which are sensitive to the specific kind of anomalous diffusion model; however, this will likely require a more extensive training. 
Another interesting extension would be to train a network on data of the type mentioned above which is characterized by different exponents on different time scales to infer the whole profile of the MSD as a function of time instead of separately considering its distinct diffusive regimes. This extension may not be completely trivial since one would have to learn more parameters that can be used to parametrize more general curves.

\begin{acknowledgements}
SB thanks Matthias Geilhufe, Bart Olsthoorn, and Aykut Argun  for inspiring discussions. FS thanks Sabareesh K. P. Velu for his supervision of the speckles experiments and fruitful discussions.
GV and SB thank Juan Ruben Gomez-Solano for inspiring discussions.
SB and RE acknowledge financial support from the Swedish Research Council (Vetenskapsr{\aa}det) under the grants No.~638-2013-9243 and No.~2016-05412.
FS and GV acknowledge financial support from H2020 European Research Council (ERC) Starting Grant ComplexSwimmers (677511).
\end{acknowledgements}

%\bibliography{learn_anom}{}

%\bibliographystyle{apsrev4-2}

\end{document}